\documentclass[twocolumn,aps,prd,preprintnumbers,nofootinbib]{revtex4-1}
\usepackage{graphicx}
\usepackage{amsmath}
\usepackage{amssymb}
\usepackage{amsfonts}
\usepackage{hyperref}
\usepackage{url}
\usepackage{color}
\usepackage{bm}
\usepackage[dvipsnames]{xcolor}

\newcommand{\be}{\begin{equation}}
\newcommand{\ee}{\end{equation}}
\newcommand{\ba}{\begin{eqnarray}}
\newcommand{\ea}{\end{eqnarray}}
\newcommand{\nn}{\nonumber}

\renewcommand{\[}{\begin{equation}}
\renewcommand{\]}{\end{equation}}

\def\lcdm{$\Lambda$CDM }

\usepackage{array}
\makeatother

\begin{document}

\preprint{IFT-UAM/CSIC-24-133}

%\title{A swampland conjecture DESIder\'atum?}
\title{Reconstruction of the swampland conjectures with DESI DR1 BAO data}

\author{Rub\'{e}n Arjona}
\email{ruben.arjoname@gmail.com}

\author{Savvas Nesseris}
\email{savvas.nesseris@csic.es}

\affiliation{Instituto de F\'isica Te\'orica UAM-CSIC, Universidad Auton\'oma de Madrid,
Cantoblanco, 28049 Madrid, Spain}

\date{\today}

\begin{abstract}
The swampland conjectures (SCs) propose constraints on effective field theories that can arise from a consistent theory of quantum gravity. Two prominent SCs suggest that the scalar field excursion and the gradient of the potential should be at most $\mathcal{O}(1)$ in Planck units. Using the first data release from the Dark Energy Spectroscopic Instrument (DESI) survey and model-independent reconstructions of the SC-related quantities at late times, via a machine learning approach known as genetic algorithms, we evaluate the consistency of these reconstructions with the SC expectations. Our results indicate that the reconstructed second SC is several sigmas away from zero, suggesting a steep potential, in contrast to recent model-specific analyses assuming exponential potentials. The novelty of our approach lies in using solely model-independent reconstructions of cosmological observables from DESI, such as the angular diameter distance and the Hubble expansion history. This makes our method readily applicable to forthcoming data from stage IV surveys, providing a framework for further assessing consistency with SCs.
\end{abstract}
\maketitle

% ------------------------------------------------------------------
\textit{Introduction:} Over the past two decades, observations using type Ia supernovae (SnIa), cosmic microwave background (CMB) anisotropies, and large-scale structure data have suggested the existence of a repulsive force acting on cosmological (large) scales, causing the Universe to undergo a phase of accelerated expansion at late times. Within the framework of general relativity (GR), this behavior implies the presence of a form of matter  with a negative equation of state parameter $w$, known as dark energy (DE). 

Currently, the most plausible phenomenological explanation for this accelerated expansion, based on available data \cite{Aghanim:2018eyx}, is the cosmological constant $\Lambda$, which acts as a uniform vacuum energy pervading all space and cold dark matter (CDM) model, collectively known as the $\Lambda$CDM model. Despite its success with observations, $\Lambda$CDM also faces some challenges, see Refs.~\cite{Perivolaropoulos:2021jda, Efstathiou:2024dvn} for recent reviews, but also some theoretical problems \cite{Martin:2012bt}.

Specifically, the cosmological constant $\Lambda$ poses significant fine-tuning issues, as its value is much smaller than what quantum field theories predict \cite{Weinberg:1988cp,Carroll:2000fy,Martin:2012bt}. This has led to the development of numerous alternative DE models. These models include ad-hoc ideal fluids and new scalar fields that mediate forces between particles. Examples include canonical scalar fields \cite{Ratra:1987rm,Wetterich:1987fm,Caldwell:1997ii}, scalar fields with generalized kinetic terms \cite{chiba2000kinetically,ArmendarizPicon:2000dh,ArmendarizPicon:2000ah}, models with non-minimal couplings \cite{chiba1999quintessence,Uzan:1999ch,Perrotta:1999am,Riazuelo:2001mg}, and coupled DE models \cite{Dent:2008vd}, all within or extending GR.

One of the simplest approaches to describing the background dynamics using GR is to model dynamical DE with a single minimally coupled scalar field. Typically, this involves using a field potential with a local minimum at a positive value, resulting in a stable or meta-stable de Sitter (dS) vacuum and yielding a positive vacuum energy. Another noteworthy scenario is found in quintessence models, where the potential remains positive, but the scalar field is not at a minimum, see Refs.~\cite{Amendola:2015ksp} or \cite{Tsujikawa:2013fta} for a detailed review. This situation can arise when the magnitude of the gradient of the potential $\left|V' \right |$ is sufficiently small and comparable to the potential $V$.

A prominent candidate for DE is quintessence, characterized by a single slowly rolling minimally coupled scalar field that results in accelerated expansion \cite{Amendola:2015ksp}. This scalar field can influence both the early and late stages of the Universe by dominating its energy density, and acting as a source of DE, see Ref.~\cite{Tsujikawa:2013fta} for a review.

Another promising alternative to DE models is found in covariant modifications of GR, known as modified gravity (MG) theories. These theories are inspired by high-energy physics, including quantum gravity and string theory, and have distinct features. In this context, GR is viewed as an effective low-energy theory that requires higher-order corrections as the energy scale increases \cite{tHooft:1974toh}. To address the plethora of DE and MG models, efforts have been made to develop a unified framework, such as the effective field theory (EFT) approach \cite{Gubitosi:2012hu,Hu:2013twa} or the effective fluid approach  \cite{Arjona:2018jhh,Arjona:2019rfn,Arjona:2020gtm,Cardona:2020ama}.

For nearly a century, theoretical physicists have sought to create a theory of quantum gravity that integrates the principles of Einstein's GR with those of quantum field theory. Although GR has shown remarkable predictive accuracy at scales below Planck length, its quantization poses significant challenges, as it is renormalizable only at one loop \cite{Birrell:1982ix}. Thus, GR may represent the low-energy limit of a more fundamental, high-energy theory.

A possible viable theory of quantum gravity is string theory, which presents a vast ``landscape" with nearly $10^{500}$ vacua, where a consistent quantum theory of gravity is believed to be formulated along with consistent low-energy EFTs. However, these ``landscapes", which can potentially arise within UV-complete quantum gravity theories, are known to be bordered by even larger regions, referred to as ``swamplands", where EFTs that appear consistent and are coupled to gravity are, in reality, inconsistent with the quantum theory of gravity \cite{Obied:2018sgi}. 

Therefore, it is essential for consistent EFTs to avoid these ``swamplands", which has led to the development of various conjectures, such as the weak gravity conjecture proposed a decade ago and, more recently, the ``swampland conjecture" which aims to exclude any (meta-)stable de Sitter configurations within string landscapes. This makes it challenging to incorporate accelerating phases, such as DE domination and the inflationary epoch, into cosmological models \cite{das2019note, Kinney:2018nny}. 

The two proposed swampland conjectures that we will consider (which we will define as SC1 and SC2) refer to the constraints on the allowed field range of a scalar field $\phi$ defined by an EFT and to the slope of the potential of such fields respectively. In reduced Planck units these conjectures are defined as follows 
\begin{enumerate}
  \item SC1, known as the swampland distance conjecture, states that when a scalar field traverses a trans-Planckian distance in field space, a tower of states descends and becomes light, signaling a breakdown of the effective field theory. In the past literature, this has been written as: $ |\Delta \phi|\, M_\textrm{pl}^{-1}<\Delta \sim \mathcal{O}(1)$ \cite{Ooguri:2006in}.
  \item SC2, the de Sitter conjecture, suggests a lower bound for the gradient of the scalar potential,  $M_\textrm{pl}\left|\nabla_{\phi} V\right| / V>c \sim \mathcal{O}(1)$, in any consistent theory of gravity when $V>0$ \cite{Obied:2018sgi}. However, this formulation has been challenged by explicit counterexamples \cite{denef2018sitter}, leading to a refined version that also imposes constraints on the second derivative of the potential.
\end{enumerate}

\noindent Here, $\Delta$ and $c$ are positive constants of order 1 and the reduced Planck mass is $M_\textrm{pl}=1/\sqrt{8\pi G}$. 
The second swampland criterion SC2 is driven by the difficulty in reliably constructing de Sitter vacua and the insights gained from string theory models of scalar potentials \cite{agrawal2018cosmological}. It is not satisfied in the $\Lambda$\text{CDM} model because having a positive cosmological constant or being at the minimum of a potential with positive energy density breaches the bound \cite{Agrawal:2018own}. Therefore, a quintessence model, which involves a rolling scalar field potential, would be necessary. Consequently, if our model-independent reconstructions suggest consistency with the second swampland conjecture, this could hint at possible deviations from the $\Lambda$\text{CDM} model.

The aim of the swampland conjecture (SC) is to identify constructions compatible with a quantum theory of gravity. Notably, certain quintessence models have been shown to satisfy these conjectures at late times \cite{Agrawal:2018own}. In Ref.~\cite{Elizalde:2018dvw}, the authors utilized Gaussian Processes to reconstruct the potential form from $H(z)$ data, uncovering evidence that may challenge the swampland conjecture. A similar analysis was carried out in \cite{Yang:2020jze}. In Ref.~\cite{arjona2021machine} compatible reconstructions of the SC where found with observations at low redshift through Genetic Algorithms (GAs) and cosmography reconstructions on the Hubble parameter. 

According to Refs.~\cite{Colgain:2019joh,Banerjee:2020xcn}, quintessence models and current data tend to favor a lower value of $H_0$ compared to the $\Lambda$\text{CDM} model, offering a robust consistency check of the swampland conjectures. Conversely, other studies have indicated that string-inspired quintessence models with exponential potentials are not supported by observations, and that the swampland conjectures conflict with viable single-field quintessence models \cite{Akrami:2018ylq,Raveri:2018ddi}. However, it has been suggested that this issue might be resolved with multi-field models \cite{Garg:2018reu,Akrami:2020zfz, Payeur:2024kyy}. For further insights into the implications of the swampland conjectures on DE, see  Ref.~\cite{schoneberg2023news} for a recent summary. 

However, it is clear that a top-bottom approach of just picking one specific model, e.g. in the form of a choice for potential of a one or more scalar fields, and trying to do a full numerical analysis, comparing to the data and trying to find the best one, is clearly an exercise in futility as there are practically infinite potential and infinite models. For example the well known \textit{Encyclop\ae dia Inflationaris} of Ref.~\cite{Martin:2013tda}, provides an impressive in length and scope overview of inflationary models, albeit one may always find one which is not in the list. Thus, a more bottom-up approach might be desirable in order to obtain a sufficiently broad and robust conclusion. 

In this work we will use a specific machine learning (ML) technique called the GAs, which is a  stochastic optimization approach aiming to remove model bias that can bias conclusions drawn from the data about fundamental physics. ML algorithms help eliminate biases associated with pre-selecting a specific model, making them particularly suitable for investigating poorly understood phenomena such as DE, DM, or MG. Additionally, ML allows us to reconstruct data with the least of assumptions such as adhering to a particular dark energy model. Focusing on quintessence as a case study, we use the latest compilation of the Dark Energy Spectroscopic Instrument (DESI) baryon acoustic oscillations (BAO) data release (DR1) data \cite{adame2024desi} to analyze the cosmological implications for the two SC. Recently, there have been many analyses using the DESI data to test scalar-field models \cite{Wolf:2024eph,Ramadan:2024kmn,Bhattacharya:2024hep,Yin:2024hba,Tada:2024znt}, other generic DE models with time varying equation of state parameters \cite{Taule:2024bot,Ghedini:2024mdu,Reboucas:2024smm,Lohakare:2024ize,Dinda:2024ktd,Giare:2024gpk,Cortes:2024lgw,DESI:2024kob,Roy:2024kni,Pourojaghi:2024tmw,Chudaykin:2024gol, Park:2024jns, Wang:2024dka}, but also tests of the data themselves, e.g. litmus tests \cite{LHuillier:2024rmp,Dinda:2024kjf}, or tests on the statistical methodology \cite{Patel:2024odo}.  

\textit{Theory:}
At late times, the Friedmann equations, assuming flatness and ignoring radiation and neutrinos, can be expressed as follows:
\ba
H^{2} &=&\frac{8\pi G}{3} \left(\rho_{m}+\frac{1}{2} \dot{\phi}^{2}+V(\phi)\right), \\
\dot{H} &=&-4 \pi G\left(\rho_{m}+\dot{\phi}^{2}\right),
\ea
where the Hubble parameter is defined as $H\equiv\frac{\dot{a}}{a}$ and the scale factor is related to the redshift via $a=\frac1{1+z}$. We can now solve the previous equations for the potential and the kinetic term, so as to write them as \cite{Sahni:2006pa}
\ba
\frac{8 \pi G}{3 H_{0}^{2}} V(x) &=&\frac{H(x)^{2}}{H_{0}^{2}}-\frac{x}{6 H_{0}^{2}} \frac{d (H(x)^{2})}{d x}-\frac{1}{2} \Omega_{\mathrm{m},0}\, x^{3}\label{eq:Vx0},~~~~ \\
\frac{8 \pi G}{3 H_{0}^{2}}\left(\frac{d \phi}{d x}\right)^{2} &=&\frac{2}{3 H_{0}^{2} x} \frac{d \ln H}{d x}-\frac{\Omega_{\mathrm{m},0}\,x}{H^{2}}\label{eq:phix0},
\ea
where we have set $x\equiv1+z$. In order to numerically solve Eqs.~\eqref{eq:Vx0}-\eqref{eq:phix0}, it is more convenient to rescale all variables and use dimensionless quantities. This can be achieved, for instance, by introducing the Planck mass $M_\textrm{pl}\equiv\sqrt{\frac{\hbar c}{8 \pi G}}=\sqrt{\frac{1}{8 \pi G}}$ in natural units ($\hbar=c=1$) and by considering that the critical density is $\rho_c=\frac{3H_0^2}{8\pi G}$. We can then make the following redefinitions
\ba 
E(z)&\equiv&H(z)/H_0, \nn \\
\tilde{\phi}(z)&\equiv&\frac{\phi(z)}{\sqrt{3}\,M_\textrm{pl}},\nn \\
\tilde{V}(z)&\equiv&\frac{V(z)}{\rho_c},
\ea 
and reformulate the reconstruction equations for the scalar field as 
\ba
\tilde{V}(x) &=&E(x)^2-\frac{x}{6} \frac{d (E(x)^{2})}{d x}-\frac{1}{2} \Omega_{\mathrm{m},0}\,x^{3}\label{eq:Vx}, \\
\left(\frac{d \tilde{\phi}}{d x}\right)^{2} &=&\frac{2}{3x} \frac{d \ln E}{d x}-\frac{\Omega_{\mathrm{m},0}\,x}{E(x)^{2}}\label{eq:phix}.
\ea
To reconstruct the potential we need the dimensionless Hubble parameter $E(x)$ and a value for the matter density $\Omega_{\mathrm{m},0}$. We can then proceed as follows: first, we can integrate Eq.~(\ref{eq:phix}) either numerically or analytically (if possible) to determine $\tilde{\phi}(x)$ up to a constant, then we write $x$ as a function of $\tilde{\phi}$ i.e $x(\tilde{\phi})$ and insert it in  Eq.~(\ref{eq:Vx}) to find the potential in terms in the scalar field $\tilde{V}(\tilde{\phi})$.

In order to simplify the analysis later on, we can also define the following quantities related to the first and second derivatives of the potential \cite{Copeland:2006wr}:
\ba 
\lambda &\equiv& M_\textrm{pl}\frac{|V'(\phi)|}{V(\phi)}= \frac{1}{\sqrt{3}}\frac{\tilde{V}'(\tilde{\phi})}{\tilde{V}(\tilde{\phi})},\label{eq:lambda}\\
\Gamma &\equiv& \frac{V(\phi)\,V''(\phi)}{V'(\phi)^2}=\frac{\tilde{V}(\tilde{\phi})\,\tilde{V}''(\tilde{\phi})}{\tilde{V}'(\tilde{\phi})^2}.\label{eq:gamma}
\ea 
In general, $\lambda$ and $\Gamma$ will not be constants and may depend on $\phi$. Then, the SC2 is equivalent to $\lambda >c \sim \mathcal{O}(1)$, while for $\Gamma$ we must have the condition $\Gamma>1$ in order to have tracking solutions and accelerated expansion at late times \cite{Copeland:2006wr}. Similarly, the SC1 becomes in this notation $\sqrt{3}\Delta \tilde{\phi}<\Delta \sim \mathcal{O}(1) $. In what follows, we will provide reconstructions for all aforementioned quantities $(\Delta \tilde{\phi}, \lambda, \Gamma)$ and interpret what they imply for the conjectures. 

In this work we will reconstruct the aforementioned quantities, using the latest BAO DESI data, which include measurements of the angular diameter distance $d_\mathrm{A}(z)$ and the Hubble parameter, in the form of specific combinations of the two, see Ref.\cite{adame2024desi}. These two parameters though, assuming GR and the Friedmann–Lema\'itre–Robertson–Walker metric, are related via the comoving distance at some redshift $z$ as \cite{Weinberg:2008zzc} 
\be 
r(z)=\frac{c}{H_0}\frac1{\sqrt{-\Omega_{k,0}}}\sin \left(\sqrt{-\Omega_{k,0}}\int_0^z\frac{c}{H(z')/H_0}dz'\right),\label{eq:rz}
\ee 
and the angular diameter distance is related to the comoving one via
\ba
d_\mathrm{A}(z)&=& (1+z)^{-1} r(z). \label{eq:dAeq}
\ea 
Using Eq.(\ref{eq:dAeq}) and assuming flatness, we can relate the dimensionless Hubble parameter  $E(z)$ and the comoving distance $d_\mathrm{A}(z)$ as
\ba
\label{eq:HdA}
E(z)=\frac{1}{r'(z)}\left(\frac{c}{H_0}\right).
\ea
To perform the reconstruction, we find it is easier to extract from the BAO DESI data the angular diameter distance $d_\mathrm{A}(z)$ and then use Eq.~\eqref{eq:HdA} to get the dimensionless Hubble parameter. Furthermore, we will also assume a prior on the matter density parameter $\Omega_{\mathrm{m},0}$, as it appears as an undetermined constant in Eqs.~\eqref{eq:Vx} and \eqref{eq:phix}, with the value $\Omega_{\mathrm{m},0}=0.3160\pm 0.0065$ which corresponds to the $(w_0,w_a)$ best-fit combination of DESI, CMB and DESY5 data; see Table 3 in Ref.~\cite{adame2024desi}. 

It is important to note that with the aforementioned equations, one can attempt to reconstruct a plethora of DE models, such as the constant equation of state $w$ = const. model or other parameterized $w(z)$ models \cite{Scherrer:2015tra}. However, for models in which the equation of state parameter crosses $w=-1$, known as the phantom-divide line, then the kinetic term of Eq.~\eqref{eq:phix} becomes negative, corresponding to complex values of the scalar field, which is impossible as in this scenario the field is assumed to be real. In general, a single minimally coupled scalar field cannot cross the phantom-divide line $w=-1$ \cite{Nesseris:2006er}.

% ----------------------------------------------------------------------------------------------------------
%\section{Reconstruction \label{sec:data}}
%\subsection{Data \label{sec:data1}}
\textit{Reconstruction:} The data used is our work are  the DESI DR1 BAO obtained via the following target samples: the Bright Galaxy Sample $\left(\text{BGS}, 0.1<z<0.4\right)$  the Luminous Red Galaxy Sample (LRG, $0.4<z<0.6$ and $0.6<z<0.8)$, the Emission Line Galaxy Sample (ELG, $1.1<z<1.6$, the combined LRG and ELG Sample (LRG + ELG, $0.8<z<1.1)$, the Quasar Sample $($QSO, $0.8<z<2.1)$ and the Lyman-$\alpha$ Forest Sample (Ly$\alpha$ $1.77<z<4.16)$. The statistics and data points used for our reconstruction on the angular diameter distance $d_\mathrm{A}(z)$ can be found in Table 1 of Ref.~\cite{adame2024desi}. This recent DESI DR1 data indicate a preference for dynamical DE over the cosmological constant when combined with CMB and SnIa data \cite{adame2024desi}, however, this claim should be treated with care, see  Refs.~\cite{carloni2024does,colgain2024does}.

\begin{figure*}[!t]
\centering
\hspace*{-4mm}
\includegraphics[width = 0.34\textwidth]{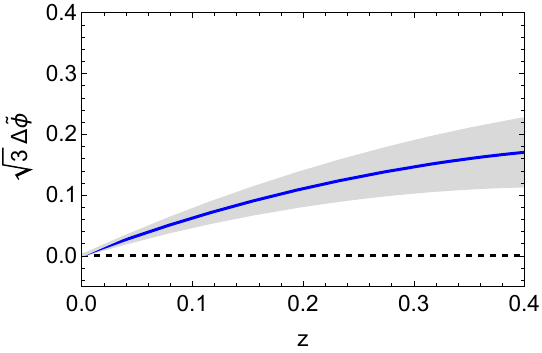}
\includegraphics[width = 0.325\textwidth]{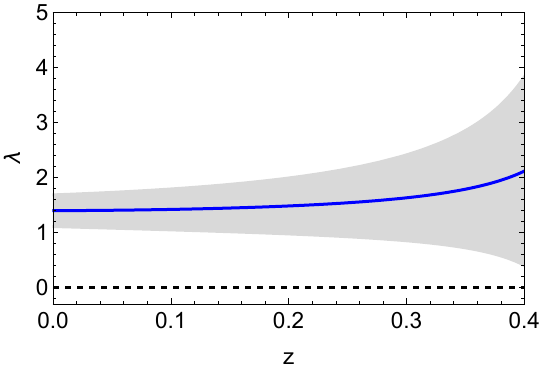}
\includegraphics[width = 0.34\textwidth]{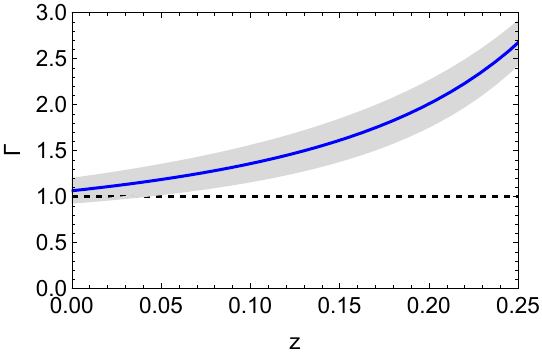}
\caption{The GA reconstruction of the first and second swampland conjectures on the left and center panels respectively, along with the $\Gamma$ parameter (right panel), which is related to the second derivative of the potential, see Eq.~\eqref{eq:gamma}. \label{fig:pot_err} }
\vspace*{-3mm}
\end{figure*}

% ---------------------
%\subsection{Genetic Algorithms \label{sec:GA}}
Here we use the GAs, which are a set of machine learning (ML) techniques designed for non-parametric data reconstruction, based on the principle of grammatical evolution and utilizing genetic operations such as crossover and mutation. For a comprehensive discussion on GA and their various cosmological applications, refer to Refs.~\cite{Bogdanos:2009ib, Nesseris:2010ep, Nesseris:2012tt, Nesseris:2013bia, Sapone:2014nna, Arjona:2020doi, Arjona:2020kco, Arjona:2019fwb, Arjona:2020axn, arjona2022testing, aizpuru2021machine, arjona2021complementary, arjona2021novel}. GAs mimic the process of evolution through natural selection, where a population of individuals evolves over time influenced by stochastic operators: mutation, a random alteration in an individual, and crossover, the combination of two or more different individuals to produce new offspring. Each individual's ``reproductive success" is usually quantified by a $\chi^2$ statistic, which is related to its fitness, indicating how well each individual in the population fits the data.

In the following paragraphs we will describe the process of reconstructing the angular diameter distance $d_\mathrm{A}(z)$ using data from DESI. First, an initial population of functions is randomly generated, with each member of the population providing initial estimates for $d_\mathrm{A}(z)$. We also apply reasonable physical priors, such as $d_\mathrm{A}(z = 0) = 0$ at the present day ($z = 0$). Importantly, at this point we do not assume any specific DE model but we do assume spatial flatness in order to express the Hubble function $H(z)$ as a function of $d_\mathrm{A}(z)$, see Eq.~\ref{eq:HdA}.

Next, the fitness of each function is evaluated using a $\chi^2$ statistic, incorporating the $d_\mathrm{A}(z)$. Subsequently, the mutation and crossover operators are stochastically applied to the best-fitting functions from each generation, which are selected via the tournament selection method (see \cite{Bogdanos:2009ib} for more details). This process is repeated thousands of times with different random seeds to ensure convergence and to avoid bias from a specific random seed choice. Once the GA code has converged, the final output is a reconstructed $d_\mathrm{A}(z)$. Note that all the DESI data points depend on the sound horizon at the drag redshift $ r_\mathrm{d} = r_\mathrm{s}(z_\mathrm{d}) $, which is challenging to estimate in a model-independent manner. Therefore, we treat $r_\mathrm{d}$ as a free parameter and determine its value from the data by minimizing the total $\chi^2$ with respect to it.

To estimate the errors on the reconstructed functions, we use an analytical approach developed in Refs.~\cite{Nesseris:2012tt, Nesseris:2013bia}. This method derives errors through a path integral over the entire functional space that the GA can explore. This path integral approach has been validated by comparison with bootstrap Monte-Carlo error estimates \cite{Nesseris:2012tt}, showing excellent agreement between the two methods.

Thus, the GA allows us to reconstruct any cosmological function, such as the $d_\mathrm{A}(z)$ considered here, by applying the algorithm to any relevant dataset. This method does not require any specific cosmological model or assumptions about DE, making our results DE model-independent. To avoid spurious reconstructions and overfitting, we conduct several runs with different random seeds and ensure that all reconstructed functions and their derivatives are continuous.
% -------------------------------------------------
%\section{Results \label{sec:results}}

\textit{Results:} Here we now present our GA reconstructions of the swampland conjectures. Specifically, in Table~\ref{tab:GAchi2} we show the best-fit $\chi^2$ values for the best-fitting GA function compared to the best-fitting \lcdm model to the same data. As can be seen, the GA achieves a lower best-fit $\chi^2$ than the \lcdm model, indicating a better performance. Note however, that our goal is not model selection per se, i.e. to find the best model in the Bayesian sense, but rather to find a formula that is a sufficiently good fit to the data (at least as good as the \lcdm model) and is based on as few assumptions as possible, so that we can interpret it in a model-independent fashion.

With this in mind, then in Fig.~\ref{fig:pot_err}, we show our GA reconstructions of the first and second swampland conjectures in the left and right panels, respectively. In each case, the blue solid line represents the GA best-fit, while the grey region indicates the $1\sigma$ error margin. As can be seen, we find that both swampland conjectures are satisfied, in contrast to what has been found thus far in the literature for example for the second conjecture in the form of $\lambda<0.6$ \cite{Alestas:2024gxe,Andriot:2024jsh,Bhattacharya:2024hep}. We also performed several tests in order to check the validity of our reconstructions under various assumptions, see Appendix~\ref{sec:app1}.

\begin{table}[]
\centering
\begin{tabular}{cccc} 
Model & $\chi^2$ & $\Omega_{\mathrm{m},0}$ \\
\hline$\Lambda \mathrm{CDM}$ & 13.60 & $0.289\pm 0.006$ \\
\hline GA & 11.85 & -   \\
\hline
\end{tabular}
\caption{The $\chi^2$ for \lcdm and GA using the DESI data.}
\label{tab:GAchi2}
\end{table}

%----------------------------------------------------------------------------------------------------------

%\section{Conclusions \label{sec:conclusions}}
\textit{Conclusions:} A wealth of observational evidence has confirmed that the Universe is undergoing accelerated expansion, with the cosmological constant $\Lambda$ being one of the simplest explanations for DE. However, a significant issue is that the observed energy scale of DE is much smaller than the energy scale predicted by vacuum energy in particle physics \cite{Weinberg:1988cp,Carroll:2000fy,Martin:2012bt}. Alternative explanations include models with slowly rolling scalar fields that govern the Universe's late-time accelerated expansion, with quintessence being one such example.

\begin{figure*}[!t]
\centering
\hspace*{-4mm}
\includegraphics[width = 0.333\textwidth]{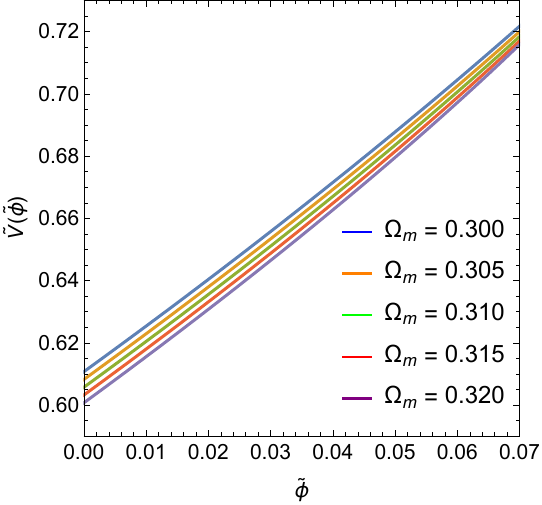}
\includegraphics[width = 0.333\textwidth]{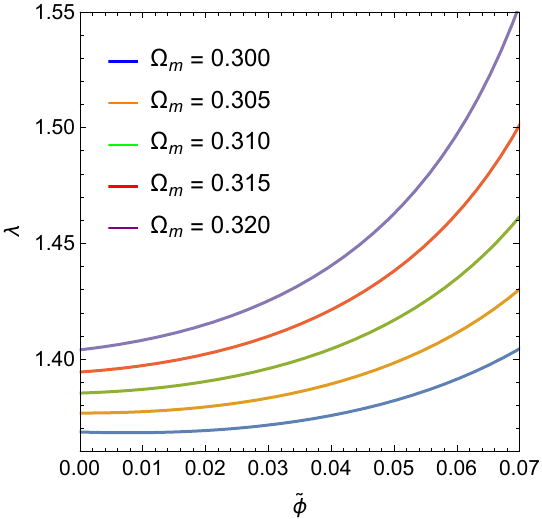}
\includegraphics[width = 0.325\textwidth]{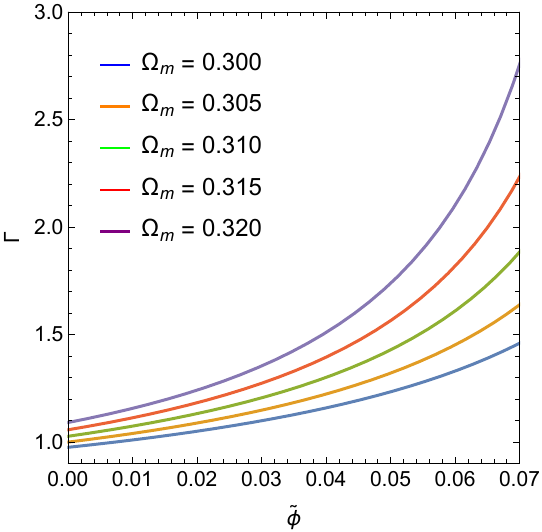}
\caption{Left panel: Shape of the dimensionless potential (see Eq.~\ref{eq:Vx}) as a function of  $\tilde{\phi}$ from our GA reconstruction with different values of the matter density $\Omega_\mathrm{m,0}$. In all cases one can see that the potential is steep. Middle panel: The GA reconstruction of the second swampland conjecture defined as $\lambda$ with different values of the matter density $\Omega_\mathrm{m,0}$ which serves to highlight that despite the reasonable value of  $\Omega_\mathrm{m,0}$ one uses, the shape maintains a similar form. Right panel: Plot of $\Gamma$ as defined in Eq.~\eqref{eq:gamma} for different values of the matter density $\Omega_\mathrm{m,0}$.\label{fig:V_phi}}
\vspace*{-1mm}
\end{figure*}

\begin{figure*}[!t]
\centering
\hspace*{-4mm}
\includegraphics[width = 0.33\textwidth]{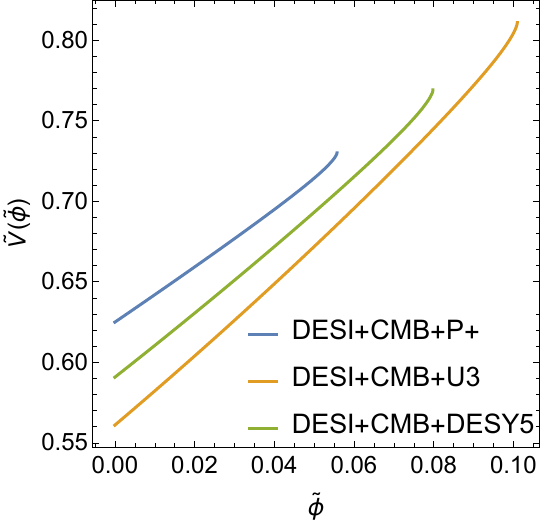}
\includegraphics[width = 0.32\textwidth]{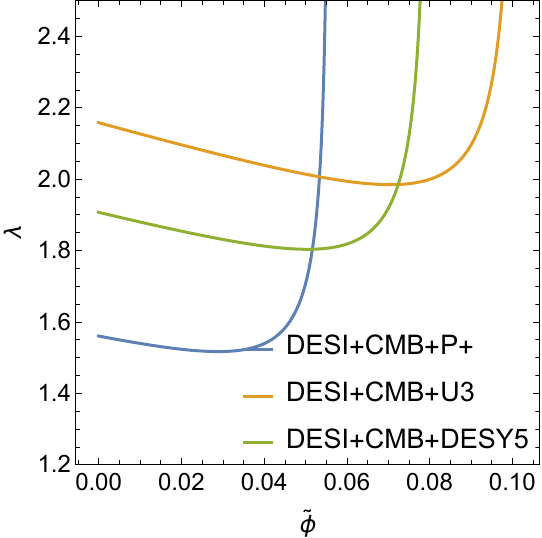}
\includegraphics[width = 0.32\textwidth]{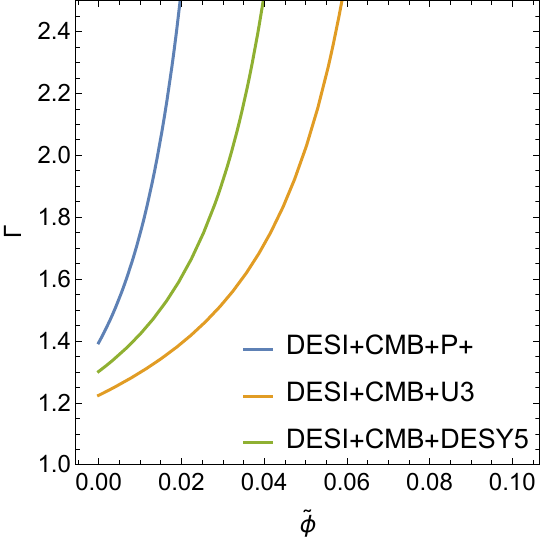}
\caption{Reconstructions of a scalar-field potential $V$ (left) and the parameters $\lambda$ (middle) and $\Gamma$ (right), based on the DESI $w_0-w_a$ best-fit. As at some point the scalar field kinetic term becomes negative, the reconstruction fails. Still, the general properties of the reconstructed parameters of the scalar field, are consistent with our GA fits.} \label{fig:DESIplots}
\vspace*{-1mm}
\end{figure*}

As part of the ongoing pursuit of a theory of quantum gravity, several swampland conjectures have been proposed within the framework of string phenomenology. These conjectures outline criteria that EFTs must meet to be consistent with quantum gravity \cite{Cicoli:2012tz} and some of these conjectures have significant implications for cosmology, such as the swampland conjecture examined here.

Using the latest BAO data from the DESI collaboration and model-independent reconstructions of the expansion history of the Universe, we assess the consistency of our reconstructed SC-related quantities with the swampland conjectures. Our results indicate that the second conjecture $\lambda=\left|\nabla_{\phi} V\right| / V>c \sim \mathcal{O}(1)$, aligns with the reconstructed data at low redshifts, as the scalar field kinetic term becomes negative. This is in contrast to recent analyses assuming an exponential potential which found that $\lambda\le 0.6$, thus highlighting the ML techniques can help avoid the theoretical model selection bias. Then, future data releases from stage IV surveys like \textit{Euclid} will help place stringent constraints on the validity of the swampland conjectures. 

We gratefully thank Y.~Akrami and M.~Montero for useful discussions. The authors also acknowledge support from the research project PID2021-123012NB-C43 and the Spanish Research Agency (Agencia Estatal de Investigaci\'on) through the Grant IFT Centro de Excelencia Severo Ochoa No CEX2020-001007-S, funded by MCIN/AEI/10.13039/501100011033.

\textit{Numerical Analysis Files}: The GA codes used by the authors in the analysis of the paper can be found at \href{https://github.com/RubenArjona}{https://github.com/RubenArjona}.\\

\begin{appendix} 
\section{Consistency tests of the reconstructions\label{sec:app1}}

In this appendix we show that the general behaviour of the potential (see Eq.~\ref{eq:Vx}) and the second swampland conjecture remains the same with different values of the matter density $\Omega_\mathrm{m,0}$. On the left panel of Fig.~\ref{fig:V_phi}) we plot the potential against the scalar field $\phi$ with $\Omega_\mathrm{m,0}=[0.300,0.320]$ where we can appreciate the steepness of the potential. On the right panel of Fig.~\ref{fig:V_phi}) we plot the SC2 as a function of $\phi$ with with $\Omega_\mathrm{m,0}=[0.300,0.320]$ where despite the reasonable value of  $\Omega_\mathrm{m,0}$ one uses, the shape maintains a similar form.

In a similar vein, in Fig.~\ref{fig:DESIplots} we show reconstructions of a scalar-field potential $V$ and the parameters $\lambda$ and $\Gamma$, based on the DESI $w_0-w_a$ best-fit. As can be seen, the general behavior is the same as in our GA reconstruction, thus hinting that the effect of a value $\lambda\sim \mathcal{O}(1)$ maybe be entirely due to the DESI data.
\end{appendix}

\bibliography{recon}

\end{document}